\documentclass[prd,preprint,superscriptaddress,showpacs,byrevtex]{revtex4}

\newcommand{\be}{\begin{equation}}
\newcommand{\ee}{\end{equation}}
\newcommand{\bea}{\begin{eqnarray}}
\newcommand{\eea}{\end{eqnarray}}
\usepackage[dvips]{graphicx}
\begin{document}
%
%\draft
%
%\tighten
%\firstfigfalse
%\twocolumn[\hsize\textwidth\columnwidth\hsize\csname@twocolumnfalse\endcsname
%
\title{Non-Zero Magnetic Screening Mass in QED and QCD \\
at One Loop Level in Non-Equilibrium}
\author{Fred Cooper} \email{fcooper@lanl.gov}
\affiliation{T-8, Theoretical Division, Los Alamos National Laboratory,
Los Alamos, NM 87545, USA}
\author{Chung-Wen Kao} \email{kao@a35.ph.man.ac.uk}
\affiliation{Theoretical Physics Group, Department of Physics and Astronomy, University of Manchester,
Manchester,M13 9PL, UK}
\author{Gouranga C. Nayak} \email{nayak@shakti.lanl.gov}
\affiliation{T-8, Theoretical Division, Los Alamos National Laboratory,
Los Alamos, NM 87545, USA }

\date{\today}
\begin{abstract} 
Using the Schwinger-Keldysh closed time path integral formalism we show that
the magnetic screening mass in QED and QCD at one loop level is 
non-zero as long as the single particle  distribution function $f(\vec{k})$ is non-isotropic, {\it i.e.}
it depends on the direction of the momentum. 
For isotropic distribution functions such as those corresponding to  thermal equilibrium
the magnetic screening mass at one loop level is found to be zero 
which is consistent with finite temperature field theory. The non-zero
magnetic screening mass in non-isotropic non-equlibrium situations
has fundamental importance in that it acts as a natural cut-off 
to remove infrared divergences in the magnetic sector. Thus it 
allows one to avoid infrared problems which previously made it difficult
to use a transport theory approach using perturbative QCD or QED scattering
kernels  to study the  thermalization
of a QED or QCD plasma. 
\end{abstract}  
\maketitle
%\narrowtext   
\newpage
\section{Introduction}
An important problem to be understood in QED and QCD plasmas 
is how an out of equilibrium plasma thermalizes.
This problem is quite important in the case of the QCD plasma produced
at the RHIC and  the LHC \cite{wang}
because the system lives for few $fm/c$ after which it hadronizes. 
As the system is very short lived  there is some doubt as to whether
the system thermalizes and it is important to know the thermalization
time scales of the various components of the plasma.  As two nuclei collide almost
at the speed of light the initial system of quarks and gluons  is highly
out of equilibrium. Only after many secondary collisions might this
parton system reach local thermal equilibrium before hadronization. Thus
there is a competition between the expansion which causes cooling and leads to the
phase transition into hadrons, and the equilibration time scales.
As all the signatures for quark-gluon plasma detection depend on the
non-equlibrium space-time evolution of the partons at RHIC and LHC
\cite{geiger}, it is essential to understand the process 
of the equilibration  of partons in conditions found
at these experiments. 

Due to the impracticality of solving directly using QCD the coupled
Schwinger Dyson equations of field theory, the alternative is to 
solve the relativistic Boltzmann equation:
\be
p^{\mu}\partial_{\mu} f(x,p) = C(x,p),
\ee
in the absence of any classical background field \cite{cooper},
in order to study the thermalization of a QED or QCD plasma. 
In the above equation
$f(x,p)$ is the single particle distribution function and
$C(x,p)$ is the Boltzmann collision term describing the secondary collisions.
The Boltzmann collision term is given by:
\bea
C(x,p)=\int 
\frac{d^3p_2}{(2\pi)^3 p_2^0} \int \frac{d^3p_3}{(2\pi)^3 p_3^0}
\int \frac{d^3p_4}{(2\pi)^3 p_4^0} 
|M(p p_2 \rightarrow p_3 p_4)|^2 ~(2\pi)^4~\delta^4(p+p_2-p_3-p_4) \nonumber \\
{[f(x,p_4) f(x,p_3) (1 \pm f(x,p_2))(1 \pm f(x,p))-f(x,p_2)f(x,p)(1 \pm 
f(x,p_4)) (1 \pm f(x,p_3))]}. 
\eea
In the above expression $M(p p_2 \rightarrow p_3 p_4)$ is the matrix
element for the $p p_2 \rightarrow p_3 p_4$ collisions,
$(1 \pm f)$ is due to Bose enhancement/Fermi suppression factor. 
Once the initial distribution function is known one can solve
this equation to study the thermalization of the system. However,
there is a severe  infrared divergence problem that must be overcome before solving
the above transport equation. The infrared divergence comes from 
the matrix element squared $ |M(p p_2 \rightarrow p_3 p_4)|^2$
when the momentum transfer: $Q^2 = (p-p_3)^2 \rightarrow 0$.
This divergence occurs at small angle scattering in perturbative QED or QCD. One way to solve this
problem is to insert ad-hoc momentum transfer cut-off values in the integrations. However,
the results are quite sensitive to the values of these ad-hoc cut-offs.
A more satisfactory solution is to use  a medium modified resummed propagator
to evaluate the scattering matrix elements squared:
$ |M(p p_2 \rightarrow p_3 p_4)|^2$ \cite{heis}. 
In this way the medium dependent
screening masses act as a natural cut-off for the infrared divergences. 
The calculation of the screening mass taking into account the contribution
of all loops  
is an extremely difficult task and one therefore limits the
calculation to the one or two-loop level. Although the Debye screening mass
has been shown to be non-zero in equilibrium at one loop order, the magnetic screening
mass is zero at one loop as shown by one loop calculations using
finite temperature field theory methods \cite{pis,bz,thoma}.
Hence a general supposition  has been that even if the electric field
is sceened the magnetic field is not screened at one loop level and thus one
has an infrared problem to contend with in that approximation.

In this paper we show that this supposition is not necessarily true. We find that the magnetic field is screened
at the one loop level as long as the medium is non-isotropic. 
Since parton model considerations  lead to non-isotropic initial conditions for the plasma,
this leads to a natural solution to the infrared problem.  All the
earlier calculations done \cite{pis,bz,thoma}
assumed either equilibrium or isotropic distribution
functions. We will show that the magnetic screening mass in QED and QCD at
one loop level is non-zero for non-isotropic non-equlibrium situations.
Hence there seems to be no natural infrared problem in  non-equilibrium
non-isotropic situations. This solves the problem of having infrared divergences
in transport approaches which study 
 the thermalization  of non-equilibrium QED and QCD plasmas.  Of course one expects that 
an exact study would provide non-perturbative effects of similar magnitude to the contribution
to the magnetic mass due to non-isotropic effects.  These effects are beyond the scope
of this paper. 

The paper is organized as follows: In section II we derive the one loop
magnetic and Debye screening mass formula in QED and QCD by using the
closed-time path  (CTP) formalism
\cite{clos}. We summarize and conclude our main results in section III.

\section{Magnetic Screening Mass in 
QED and QCD at One loop Level}
In this section we will derive the magnetic and Debye screening mass
in QED and QCD at the one loop level. In case of QCD
we will determine only the quark loop
contribution in this paper.  The gluon loop contribution to these
screening masses was discussed in a  previous paper \cite{fred} and we will combine
these two results to discuss the total one loop contribution later. As we are
interested in  
the infrared properties we will consider the infrared limit to 
define magnetic and Debye screening masses \cite{rebhan}: {\it i.e.} we will
take 
\bea
m_D^2 ~= ~Re\Pi^L(p_0 =0, |\vec p| \rightarrow 0) ~~~~~~~{\rm and} \nonumber \\
m_g^2 ~= ~Re\Pi^T(p_0 =0, |\vec p| \rightarrow 0), 
\label{dg}
\eea
where $\Pi^L, \Pi^T$ are the longitudinal and transverse component
of the self energy as defined below and $Re$ means the real part of the
self energy. The imaginary part of the self energy gives the
damping rates \cite{pis,thoma}.

\subsection{Magnetic Screening Mass in QED at One Loop Level} 

The diagram for the photon self energy in QED at one loop level
is shown in Fig. 1. The retarded self energy part can be written as
\cite{thoma}:
\bea
\Pi^{\mu\nu}_R(p) ~=~ie^2 \int \frac{d^4k}{(2\pi)^4}Tr[\gamma^{\mu}
S_{11}(k-p) \gamma^{\nu} S_{11}(k)~-~\gamma^{\mu}S_{21}(k-p) \gamma^{\nu}
S_{12}(k)],
\label{qs1}
\eea
where $1,2$ correspond to $+,-$ contour of the closed-time path.
\begin{figure}
   \centering
   \includegraphics{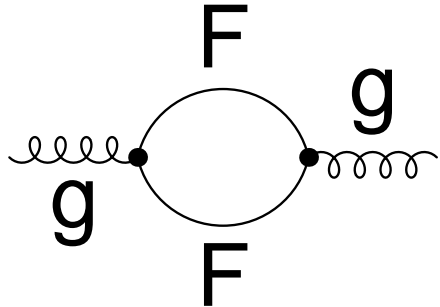}
   \caption{One Loop Graphs for the Gluon (Photon) 
Self-Energy With Quark (Electron) Loop.}
\label{photon}
\end{figure}

In this paper we will consider massless fermions since the infrared divergence
problems
occur in the massless limit.
The fermion propagators $S$ can be written as
$S_{ij}(k)= {k}\!\!\!\slash ~\Delta_{ij}(k)$ where
$\Delta_{ij}(k)$ are defined below. 
Expressing the results in terms of retarded, advanced and symmetric 
propagators we get from the above equation:
\bea
\Pi^{\mu\nu}_R(p) ~=~\frac{ie^2}{2} \int \frac{d^4k}{(2\pi)^4}Tr[\gamma^{\mu} 
({k}\!\!\!\slash - {p}\!\!\!\slash)\gamma^{\nu} {k}\!\!\!\slash][ \Delta_{S}
(k-p) \Delta_{R}(k)~+~\Delta_{A}(k-p) \Delta_{S}(k) \nonumber \\
~+~\Delta_{A}(k-p) \Delta_{A}(k) ~+~\Delta_{R}(k-p) \Delta_{R}(k)],
\label{qs2}
\eea
where
\bea
\Delta_{\stackrel{R}{A}}(k) = \frac{1}{k^2\pm i ~sgn(k_0) \epsilon}  ~~~~~~~~{\rm and}~~~~~~~
\nonumber \\
\Delta_S(k) = -2\pi i \delta(k^2) [1- 2\theta(k_0) f(\vec k) -2 \theta(-k_0)
f(-\vec k)].
\label{prop}
\eea
In the above equations
$f(\vec k)$ is the non-equilibrium distribution function which must satisfy the constraint of
yielding a finite number and energy density upon integration. As we are interested in studying
the properties of the medium we will neglect the vacuum part of the
self energy from the above equation. The divergence in the vacuum part of the
self energy is absorbed in the redefinition of the bare quantities.
 We consider here only the terms which are
proportional to the distribution function: we will 
not consider the combination $\Delta_R \Delta_R$ and
$\Delta_A \Delta_A$ in eq. (\ref{qs2}). 
After performing the trace in eq. (\ref{qs2}) we get:
\bea
\Pi^{\mu\nu}_R(p) ~=~2ie^2 \int \frac{d^4k}{(2\pi)^4}[(k-p)^{\mu} k^{\nu} 
+k^{\mu} (k-p)^{\nu} -g^{\mu \nu} (k-p)\cdot k] \nonumber \\
\cdot{[ \Delta_{S}
(k-p) \Delta_{R}(k)~+~\Delta_{A}(k-p) \Delta_{S}(k)]}.
\label{qs3}
\eea
In the rest frame of the medium the longitudinal and transverse components of 
the self energies are given by \cite{cw,thoma}:
\be
\Pi^L(p)=L_{00}(p)\Pi^{00}(p) ~~~~~~~{\rm with}~~~~~~~L_{00}(p)=-
\frac{|\vec p|^2}{p^2};
\label{pil}
\ee
and
\be
\Pi^T(p)=\frac{1}{2}T_{ij}(p)\Pi^{ij}(p)~~~~~~~~~~{\rm with}~~~~~~~~
T_{ij}(p)=-\delta_{ij} +\frac{p_i p_j}{|\vec p|^2},
\label{pit}
\ee
where we have assumed that we are interested in a static limit ($p_0$ =0 and
$|\vec p| \rightarrow 0$) so that we have contracted only with $L_{00}$
and $T_{ij}$ in the above. If we do not consider the static limit we have
to consider more combinations in the above equations \cite{cw}. 

Let us first consider the transverse component
$\Pi^T_R(p)$ of the self energy:
\bea
&&\Pi^T_R(p) ~= ~\frac{1}{2}(-\delta_{ij} 
+\frac{p_i p_j}{|\vec p|^2})\Pi^{ij}_R(p)~=~ 
~ie^2 \int \frac{d^4k}{(2\pi)^4}[
(-\delta_{ij} +\frac{p_i p_j}{|\vec p|^2}) \nonumber \\
&&((k-p)^i k^j +k^i (k-p)^j -g^{ij} (k-p)\cdot k)] 
{[ \Delta_{S} (k-p) \Delta_{R}(k)~+~\Delta_{A}(k-p) \Delta_{S}(k)]}.
\label{pt1}
\eea
Simplifying the above equation we get:
\bea
\Pi^T_R(p) ~=
~-2ie^2 \int \frac{d^4k}{(2\pi)^4}&&[(k_0-p_0)k_0-(\vec k \cdot \hat p)
(\hat p \cdot (\vec k-\vec p))]
\nonumber \\
&&{[ \Delta_{S} (k-p) \Delta_{R}(k)~+~\Delta_{A}(k-p) \Delta_{S}(k)]}.
\label{pt2} 
\eea
Using the expressions for $\Delta$ from eq. (\ref{prop}) and
taking the real part we find:
\bea
Re \Pi^T_R(p)~=~
~-2ie^2 \int \frac{d^4k}{(2\pi)^4}[(k_0-p_0)k_0-(\vec k \cdot \hat p)
(\hat p \cdot (\vec k-\vec p))] \nonumber \\
\{[\frac{-2\pi i \delta((k-p)^2)}{k^2} 
[1- 2\theta(k_0-p_0) f(\vec k-\vec p) -2 
\theta(-k_0+p_0) f(-\vec k+\vec p)]~ \nonumber \\
+\frac{-2\pi i \delta(k^2)}{(k-p)^2} 
[1- 2\theta(k_0) f(\vec k) -2 \theta(-k_0) f(-\vec k)]\}.
\label{pt3}
\eea
Setting $p_0$= 0 and again dropping the terms which does not involve the
distribution function (the vacuum contribution) we get from the
above equation:
\bea
Re \Pi^T_R(p_0=0, \vec p)~=~
~8\pi e^2 \int \frac{d^4k}{(2\pi)^4}[(k_0^2-(\vec k \cdot \hat p)
(\hat p \cdot (\vec k-\vec p))] \nonumber \\
\{[\frac{ \delta(k_0^2-|\vec k -\vec p|^2)}{k_0^2-|\vec k|^2} 
[\theta(k_0) f(\vec k-\vec p) + \theta(-k_0) f(-\vec k+\vec p)]~ \nonumber \\
+~ \frac{\delta(k_0^2-|\vec k|^2)}{(k_0^2-|\vec k - \vec p|^2)} 
[\theta(k_0) f(\vec k) +\theta(-k_0) f(-\vec k))]\}.
\label{pt4}
\eea
More simplification can be done by using the $\delta$ function 
integration which yield:
\bea
Re \Pi^T_R(p_0=0, \vec p)~=&&\frac{4\pi e^2}{(2\pi)^4}
 \int \frac {d^3k} {|\vec k|^2 - |\vec k-\vec p|^2}            \nonumber \\
&&\{[-|\vec k-\vec p|^2+
(\vec k \cdot \hat p) (\hat p \cdot (\vec k-\vec p))] 
\left(\frac{f(\vec k-\vec p) +f(-\vec k+\vec p)}{|\vec k-\vec p|}\right)~ \nonumber \\
&&+[|\vec k|^2-(\vec k \cdot \hat p) (\hat p \cdot (\vec k-\vec p))] 
\left(\frac{f(\vec k) +f(-\vec k)}{|\vec k|}\right)]\}.
\label{pt5}
\eea
In the limit $|\vec p| \rightarrow 0$ we expand: $f(\vec k - \vec p)$ 
and $\frac{1}{|\vec k -\vec p|}$ to the leading order and use: 
 $f(\vec k - \vec p) \simeq f(\vec k) -
\vec p \cdot \nabla_{k} f(\vec k)$ and $\frac{1}{|\vec k -\vec p|} \simeq 
\frac{1}{|\vec k|}[1+\frac{\vec p \cdot \vec k}{|\vec k|^2}]$ in the
rest of our calculation. This yields:
\bea
&&Re \Pi^T_R(p_0=0, \vec p)~\simeq\frac{4\pi e^2}{(2\pi)^4}
\int \frac{d^3k}{|\vec k|}
\frac{1}{2\vec k \cdot \vec p -|\vec p|^2}\{(2\vec p \cdot \vec k -|\vec p|^2)
[f(\vec k)+f(-\vec k)]+\nonumber \\
&&[-|\vec k|^2-|\vec p|^2+2 \vec p 
\cdot \vec k +(\vec k \cdot \hat p)^2 -\vec p \cdot \vec k] 
\frac{\vec p \cdot \vec k}{|\vec k|^2}[f(\vec k)+f(-\vec k)]- \nonumber \\
&&(1+\frac{\vec p \cdot \vec k}{|\vec k|^2})
[-|\vec k|^2-|\vec p|^2+2 \vec p 
\cdot \vec k +(\vec k \cdot \hat p)^2 -\vec p \cdot \vec k] 
[ \vec p \cdot \nabla_{k}f(\vec k)+
\vec p \cdot \nabla_{k}f(-\vec k)]\}.~~~~~~~
\label{pt6}
\eea
Taking the limit $|\vec p| \rightarrow 0$ we get:
\bea
&&m_g^2(\hat p)~=~Re \Pi^T_R(p_0=0, 
|\vec p| \rightarrow 0, \hat p)~=\nonumber \\
&&e^2 \int \frac{d^3k}{(2\pi)^3}\left[
[1 +(\hat p \cdot \hat k)^2]
[\frac{f(\vec k)+f(-\vec k)}{|\vec k|}] 
~+~ [1 -(\hat p \cdot \hat k)^2]
[\frac{\hat p \cdot \nabla_{k}}{\hat p \cdot \hat k}f(\vec k)
+\frac{\hat p \cdot \nabla_{k}}{\hat p \cdot \hat k}f(-\vec k)]\right].~~~~~~~~~~
\label{ptf}
\eea
After changing $\vec k \rightarrow -\vec k$ in the $f(-\vec k)$ part of the 
integration we get:
\bea
m_g^2(\hat p)~= 
&&2e^2 \int \frac{d^3k}{(2\pi)^3}\left[
[1 +(\hat p \cdot \hat k)^2]
\frac{f(\vec k)}{|\vec k|} 
~+~ [1 -(\hat p \cdot \hat k)^2]
[\frac{\hat p \cdot \nabla_{k}}{\hat p \cdot \hat k}f(\vec k)]\right].~~~~~~~~~~
\label{mgef}
\eea
This is the expression for the magnetic screening mass in QED 
at one loop level in terms of the non-isotropic non-equilibrium
distribution function $f(\vec k)$. For any equilibrium distribution
function $f(\vec k)=f_{eq}(k_0)$ we get from the above equation
$m_g^2=0$ which is consistent with finite temperature QED results.
It can also be checked that for any isotropic distribution 
function $f(|\vec k|)$ we also get $m_g^2=0$. Only for a non
isotropic distribution function $f(\vec k)$ do we get a non-zero
magnetic screening mass in QED at one loop level.

\subsection{Magnetic Screening Mass in QCD at One Loop Level}

First of all consider the situation of  the quark loop contribution to the
gluon self energy (see Fig. 1).
The only difference between the contribution to the  gluon self energy from a  quark loop and
the photon self energy with an electron loop is a factor of $\frac{\delta_{ab}}{2}$
which comes from $Tr[T^aT^b]$ associated with the quark-gluon vertex.
Hence the quark loop contribution to the magnetic mass coming from the
gluon self energy is given by:

\bea
m_g^2(\hat p)~= 
&&~g^2 \int \frac{d^3k}{(2\pi)^3}\left[
[1 +(\hat p \cdot \hat k)^2]
\frac{f_q(\vec k)}{|\vec k|} 
~+~ [1 -(\hat p \cdot \hat k)^2]
[\frac{\hat p \cdot \nabla_{k}}{\hat p \cdot \hat k}f_q(\vec k)]\right]~~~~~~~~~~
\label{mgqf}
\eea
where $f_q(\vec k)$ is the quark distribution function.
This expression is similar to the expression found in our earlier work
\cite{fred} where we have considered the gluon loop contributions (see Fig. 2)
in covariant gauge in the frozen ghost formalism \cite{land,petr,cw}. 
In the frozen ghost formalism \cite{land,petr,cw}, the gluon self energy
is obtained from the gluon loop and tadpole loop as shown in Fig. 2.
In this formalism the ghost does not contribute to the medium effect
because the initial density of states are chosen to be that of the
physical gluons. It can be seen that apart from color factors 
the expression for the magnetic screening mass 
of quark and gluon loops are the same.  (Here we are only considering the 
contributions proportional to the single particle distribution function.
In the vacuum sector, the quark and gluon loops contributions have opposite signs). 

Using Eq. (\ref{mgqf}) and summing over all the quark flavors and then
adding the gluon loop contributions from our previous work \cite{fred}
we obtain the final expresion for the magnetic screening mass in QCD:
\bea
m_g^2(\hat p)~= 
&&~g^2 \int \frac{d^3k}{(2\pi)^3}\{
[1 +(\hat p \cdot \hat k)^2]
\left[\frac{N_{c}f_g(\vec k)+
\sum_q f_q(\vec k)}{|\vec k|}\right] \nonumber \\
&&~+~ [1 -(\hat p \cdot \hat k)^2]
\left[\frac{\hat p \cdot \nabla_{k}}{\hat p \cdot \hat k}
[N_{c}f_g(\vec k)+\sum_q f_q(\vec k)]\right]\}.~~~~~~~~~
\label{mgqgf}
\eea
This is the expression for the magnetic screening mass in QCD at one loop
level in non-equilibrium. For any equilibrium distribution
function $f(\vec k)=f_{eq}(k_0)$ we get from the above equation
$m_g^2=0$ which is consistent with finite temperature QCD results.
It can also be checked that for any isotropic distribution 
function $f(|\vec k|)$ we also get $m_g^2=0$. 
Only for a non isotropic distribution function $f(\vec k)$ do we get a non-zero
magnetic screening mass in QCD at the one loop level.

\begin{figure}
   \centering
   \includegraphics{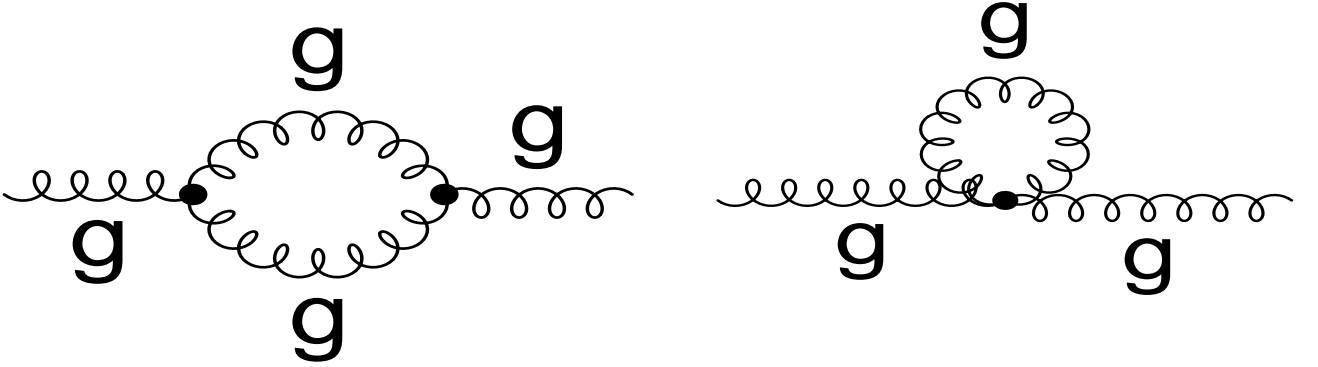}
   \caption{One Loop Graphs for the Gluon Self-Energy with Gluon Loop and 
Tadpole Diagrams}
\label{gluon}
\end{figure}

\subsection{Debye Screening Mass in QED at One Loop Level}

Taking the ($00$) component of the self energy in eq. (\ref{qs2}), using
eqs. (\ref{pil}), (\ref{dg}) and carrying out similar algebra
as above we obtain the final expression for the Debye screening mass:
\bea
m_D^2(\hat p)~=&&- 
4e^2 \int \frac{d^3k}{(2\pi)^3}
\frac{\hat p \cdot \nabla_{k}}{\hat p \cdot \hat k}f(\vec k).
\label{mdef}
\eea
This is the expression for the Debye screening mass in non-equilibrium
at the one loop level in QED 
in terms of any arbitary non-isotropic distribution function $f(\vec k)$
of the electron. For any isotropic distribution function $f(|\vec k|)$
we get from the above equation:
\be
m_D^2(\hat p)~=~
\frac{4e^2}{\pi^2} \int dk ~kf(k),
\ee
where $k=|\vec k|$.
For an equilibrium distribution function the above equation gives:
\be
m_D^2~=\frac{e^2T^2}{3},
\ee
which is the same result in QED obtained by using finite temperature 
QED methods \cite{pis,bz,thoma}.

\subsection{Debye Screening Mass in QCD at One Loop Level}
Using the quark-gluon vertex instead of the photon-electron vertex we
obtain the following  expression for the contribution to the Debye screeing mass in QCD from
a 
quark-loop:
\bea
m_D^2(\hat p)~=&&- 
2g^2 \int \frac{d^3k}{(2\pi)^3}
\frac{\hat p \cdot \nabla_{k}}{\hat p \cdot \hat k}f_q(\vec k).
\label{mdqf}
\eea
This expression for the Debye screening mass in QCD is valid in non-equilibrium 
situations for any non-isotropic distribution function $f(\vec k)$ for a single
quark flavor. Adding the gluon
loop expression in  a covariant gauge from our earlier work \cite{fred}
and summing over all quark flavors we get:
\bea
m_D^2(\hat p)~=&&- 
2g^2 \int \frac{d^3k}{(2\pi)^3}
\frac{\hat p \cdot \nabla_{k}}{\hat p \cdot \hat k}
[N_{c}f_g(\vec k)+\sum_q f_q(\vec k)].
\label{mdqgf}
\eea
This is our final result for the Debye screening mass in QCD at the one loop
level in non-equilibrium for any non-isotropic distribution function
$f(\vec k)$. For isotropic quark and gluon distribution
functions the above equation gives:
\bea
m_D^2(\hat p)~=&& 
\frac{2g^2}{\pi^2} \int dkk
[N_{c}f_g(k)+\sum_q f_q(k)],
\eea
where $k=|\vec k|$.
For an equilibrium distribution of quarks and gluons
at high temperature the above equation gives:
\bea
m_D^2~=g^2T^2\left[\frac{N_{c}}{3}+\frac{N_f}{6}\right],
\eea
which is the same result obtained in QCD by using finite temperature
QCD techniques \cite{pis,bz,thoma}.

\section{Conclusion}

Using the Schwinger-Keldysh closed time path integral formalism we have shown that
the magnetic screening mass in QED and QCD at one loop level is 
non-zero as long as the distribution function is non-isotropic, {\it i.e.}
it depends on the direction of the momentum. 
For isotropic distribution functions such as those describing
 thermal equilibrium
the magnetic screening mass at one loop level is found to be zero which
is consistent with finite temperature field theory results in QED and
QCD. The non-zero magnetic screening mass in non-isotropic non-equlibrium 
situations has fundamental importance in that it acts as a natural cut-off 
to remove infrared divergences in the magnetic sector when we
study the thermalization
of a QED or QCD plasma using a  microscopic transport theory approach.
This is particularly important at RHIC and LHC heavy-ion collisions because
it is unlikely that partons are thermalized quickly before hadronization.
As any ad-hoc infrared cut-off put by hand change the plasma properties, using
this natural infared cutoff we avoid ad-hoc infrared cutoff methods which
are sensitive to the cutoff.  We can use
the Debye screening mass discussed above as the infrared cut-off in the
electric sector and the magnetic 
mass derived in this paper as the infrared cut-off 
in the magnetic sector. Hence for non-equilibrium QED and QCD plasma
descriptions the magnetic screening mass derived in
non-equilibrium can be used to remove infrared divergences in the
magnetic sector. 

A non-equilibrium non-perturbative calculation
of the magnetic screening mass is very difficult.
However, in \cite{fred}, we showed that the values of the magnetic screening
masses found by using a single particle
distribtution function obtained from parton model considerations for
conditions relevant at
RHIC and LHC energies was comparable to screening masses obtain from lattice
QCD equilibrium results at the corresponding temperatures.
Hence in the absence of non-equilibrium non-perturbative calculations,
one can use 
the one loop level result derived in this paper to study thermalization
of QED and QCD plasma using transport methods.

\end{document}